\documentclass[twocolumn,showpacs,preprintnumbers,amsmath,amssymb,nofootinbib]{revtex4}
\usepackage{epsfig}
\def\beq{\begin{equation}}
\def\eeq{\end{equation}}
\def\bea{\begin{eqnarray}}
\def\eea{\end{eqnarray}}
\def\bq{\begin{quote}}
\def\eq{\end{quote}}

\def\gappeq{\mathrel{\rlap {\raise.5ex\hbox{$>$}}
{\lower.5ex\hbox{$\sim$}}}}
\def\lappeq{\mathrel{\rlap{\raise.5ex\hbox{$<$}}
{\lower.5ex\hbox{$\sim$}}}}

\def\ETm{E_T \!  \! \! \! \! \! \! /~~}

\def\sw2{\sin^2 \theta_W}
\def\s132{\sin^2 \theta_{13}}

\usepackage{epsfig,amsmath,graphics}
\usepackage{multirow}
\usepackage{appendix}
\usepackage{slashed}

\newcommand{\Ntwo}{${\cal N}=2$ }
\newcommand{\None}{${\cal N}=1$ }
\newcommand{\alphadot}{{\overset{.}{\alpha}}}
\newcommand{\betadot}{{\overset{.}{\beta}}}
\newcommand{\go}{{\tilde{g}}}
\newcommand{\gop}{{\tilde{g}}^\prime}
\newcommand{\sq}{{\tilde{q}}}
\newcommand{\etap}{{\eta^\prime}}
\newcommand{\chip}{{\chi^\prime}}
\newcommand{\bmat}{\begin{pmatrix}}
\newcommand{\emat}{\end{pmatrix}}
\newcommand{\be}{\begin{equation}}
\newcommand{\ee}{\end{equation}}
\begin{document}

%\preprint{APS/123-QED}

\title{How Many Supersymmetries? }% Force line breaks with \\

\author{Matti Heikinheimo, Moshe Kellerstein and Veronica Sanz}
% \email{vsanz@yorku.ca}
\affiliation{% 
Department of Physics and Astronomy, York University, \\ Toronto ON, Canada}

%\date{\today}% 

\begin{abstract}
Supersymmetry  in the gauge sector could be realized as N=1 or N=2 Supersymmetry, but the current LHC searches assume an N=1 realization. In this paper we show that squarks could be as light as few hundreds of GeV for N=2. We also describe an experimental procedure to count the number of supersymmetries, i.e. to distinguish between N=1 and  N=2 supersymmetry, based on counting bins with different jet multiplicities and number of leptons.
\end{abstract}

\pacs{Valid PACS appear here}% PACS, the Physics and Astronomy
\maketitle

\section{Introduction}

Supersymmetry, also known as  SUSY, is a very attractive proposal for physics beyond the Standard Model. SUSY provides a mechanism to stabilize the electroweak scale, hence leading to a natural Higgs scenario. SUSY also offers an explanation for Dark Matter, and may even lead to unification, i.e., the idea that the three forces of the Standard Model are actually unified into a single force at high energies. 

SUSY's successes rely on new particles near the electroweak scale, particles which should be accessible at the Large Hadron Collider (LHC). The LHC is actively searching for SUSY, so far with negative results. Current bounds on superpartner masses already undermine the theoretical appeal of SUSY.

But experimental searches make assumptions about how SUSY is realized at low energies. In particular, they are limited to only one SUSY generator, the so-called \None SUSY scenario, also known as the Minimal Supersymmetric Standard Model (MSSM). More generally, the number of SUSYs  is something we need to determine. As we explain later, a very attractive possibility is that SUSY comes in a different version, called \Ntwo SUSY~\cite{Ntwo-old}.  

In this paper we show that the current searches are very sensitive on the assumption of \None SUSY: allowing for \Ntwo SUSY opens a new area of the parameter space where light supersymmetric particles are still allowed. We then describe a set of observables which can be used to characterize the SUSY scenario. In other words, we propose a procedure to count the number of supersymmetries. This characterization is based on the current searches of ATLAS and CMS of multijets plus missing energy, with and without leptons.

To approach the problem of counting SUSYs, we exploit the fact that \None SUSY implies that the gluino, the fermionic SUSY partner of the gluon, is a Majorana fermion, whereas \Ntwo SUSY leads to a  Dirac gluino.

The MSSM is the simplest version of SUSY, and is therefore traditionally adopted in most studies. However, there are many reasons to turn towards \Ntwo SUSY: it allows natural scenarios for Dark Matter~\cite{muless, DM} without the well-tempered tension~\cite{nimawell}; it provides interesting solutions to the flavor problem in SUSY~\cite{flavor}; it has a softer behavior than \None in terms of UV structure~\cite{supersoft}; gauge unification can be achieved~\cite{muless,supersoft,Unif}, among many interesting aspects of \Ntwo SUSY~\cite{otherad}.

%%%%%%%%%%%%%%%%%%%%%%%%%%%%%%%%%%%%
%%%%%%%%%%%%%%%%%%%%%%%%%%%%%%%%%%%%
\section{Model Building}

One can ask the question of how many SUSYs lead to a reasonable low-energy theory, and the short answer is that matter multiplets should come as \None SUSY, but \None or \Ntwo SUSY in the gauge multiplets are both viable possibilities, see Ref.~\cite{supersoft} for a discussion on the subject. 

One could try to extend the matter sector to \Ntwo SUSY, as long as the gauge sector couples non-linearly to the matter sector~\cite{supersoft}, but this is a contrived possibility. One could also try to extend the gauge sector to ${\cal N}=$ 4 or higher, but then the presence of so many extra fields give rise to Landau poles for the gauge couplings at an unreasonably low scale. One could also engineer by hand a new fermionic color octet with a Dirac mass term with the gluino, and with that reproduce the same phenomenology as described in this paper. This ad-hoc modeling would not have the advantages of \Ntwo SUSY, such as the supersoft behavior, or the Dark Matter features.

For those reasons, we are going to compare \None and \Ntwo SUSY in the gauge sector, and our comparison will rely on the Majorana or Dirac nature of the colored gauge fermion, the gluino. 

%%%%%%%%%%%%%%%%%%%%%%%%%%%%%%%%%%%%
%%%%%%%%%%%%%%%%%%%%%%%%%%%%%%%%%%%%
\section{General arguments}
\label{general}

Unless there is a large gap between the SUSY colored and electroweak states, the strong production of squarks and gluinos is the dominant production mechanism for SUSY particles, and dominates over the electroweak processes, which we neglect in the rest of the paper. 

Strong production at leading order would lead to three different states 

\begin{itemize}
\item Squark pair-production ($\tilde{q}\tilde{q}$) 

where $\tilde{q}$ refers generically to a squark or an antisquark $\tilde{q}^*$. The squark pair production process is dominated by the $t$-channel gluino exchange diagram, but in the case of same flavor squark antisquark production also the $s$-channel gluon mediated process is available. If the gluino is a Dirac particle, it cannot mediate a chirality flip. Therefore the $t$-channel exchange of the Dirac gluino forbids final states  with squarks of the same helicity, and  the production cross section of the $\tilde{q}\tilde{q}$ final state is
reduced in the Dirac case as compared to the Majorana case: only
the final states with squarks of opposite helicities are allowed, i.e
$\tilde{q}_L\tilde{q}^\prime_R$ or $\tilde{q}^*_L\tilde{q}^\prime_L$ but
not e.g. $\tilde{q}_L\tilde{q}^\prime_L$.

\item Gluino pair production ($\tilde{g}\tilde{g}$)

The gluino pair production process proceeds through the $t$-channel exchange of a squark (from a $q\bar{q}$ initial state) or a gluino (from a $gg$ initial state), or through the $s$-channel process mediated by a gluon (from a $q\bar{q}$ or $gg$ initial state). Also here the $t$-channel diagram is dominant. The cross section of the
$\tilde{g}\tilde{g}$ final state is enhanced in the Dirac model as compared to the Majorana case, because
there are more gluino-degrees of freedom to be produced.

\item Gluino-Squark pair production ($\tilde{g}\tilde{q}$)

The gluino-squark pair production process is always initiated from a $qg$ or $\bar{q}g$. This process is dominated by the $t$-channel exchange of a squark or a gluino, but also the $s$-channel quark mediated diagram is available. The cross section of the $\tilde{q}\tilde{g}$ process in the Dirac model is identical to the Majorana case. A full list of the relevant Feynman diagrams and the related cross section formulae for all the processes described here can be found for example in~\cite{Choi:2010gc}.
\end{itemize}

\begin{figure}[h!]
\centering
\includegraphics[scale=0.2]{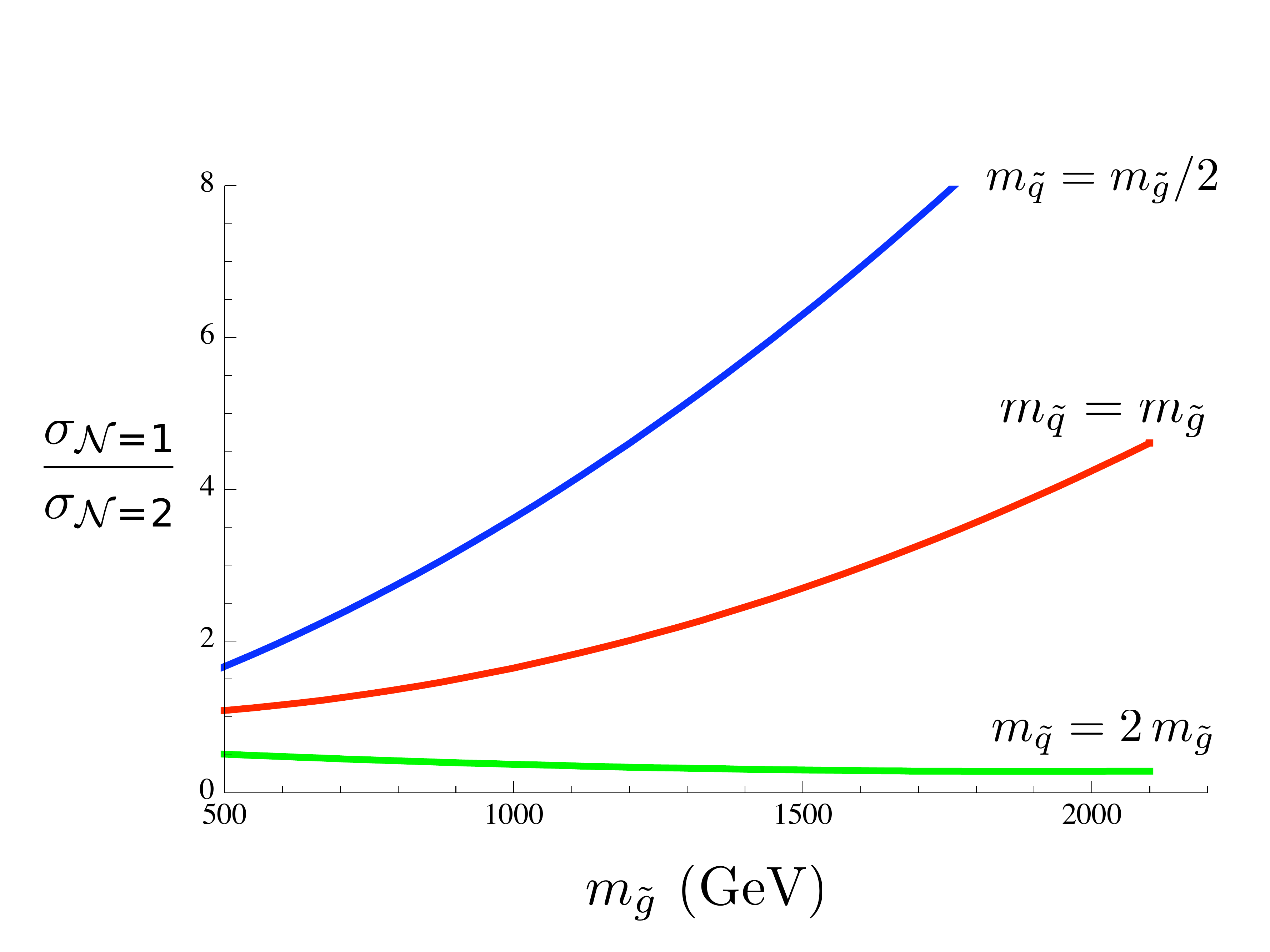}
\caption{The ratio of cross sections for  ${\cal N}=1$ and  ${\cal N}=2$, where the  total cross section is given by $\sigma=\sigma (\tilde{q} \tilde{q})+\sigma (\tilde{q} \tilde{g})+\sigma (\tilde{g} \tilde{g}) $ .}
\label{ratios}
\end{figure}

Let us then look at the total cross section of pair production of supersymmetric particles as a function of the squark and gluino masses. Obviously, the cross section of these pair production processes is different in the \None and \Ntwo cases, and the total production cross section of supersymmetric particles behaves differently as a function of the squark and gluino masses. In the low gluino mass region, where gluino pair production dominates, the total cross section is enhanced in the Dirac case with respect to the Majorana case. On the other hand, in the low squark mass region the effect is opposite. This is illustrated in Fig. \ref{ratios}, where the ratio of the production cross sections of supersymmetric particles in the MSSM and in the Dirac model is plotted as the function of the gluino mass. The squark mass is set to $\frac{1}{2}$, $1$ or $2$ times the gluino mass. As explained above, for a small squark mass, $m_{\tilde{q}}=\frac{1}{2}m_{\tilde{g}}$ the ratio grows rapidly for increasing gluino mass. This means that the overall production cross section of supersymmetric particles is heavily suppressed in that region of parameter space in the Dirac model. This will become important when looking at the current bounds from the LHC, Sec.~\ref{exclusion}.

%%%%%%%%%%%%%%%%%%%%%%%%%%%%%%%%%%%%
%%%%%%%%%%%%%%%%%%%%%%%%%%%%%%%%%%%%
\section{Simulation}
\label{simulation}

In the next section we are going to describe how we obtain bounds on \Ntwo SUSY based on the latest ATLAS search on multijets and missing energy~\cite{Aad:2011ib}. See Sec.~\ref{ATLASan} for details.
Therefore, we perform a simulation based on this experimental analysis. Following their analysis, we define a simplified model where the relevant degrees of freedom are the first generation squarks $\tilde{u}_L$, $\tilde{u}_R$, $\tilde{d}_L$ and $\tilde{d}_R$, the lightest neutralino $\chi^0$, assumed to be mostly bino, the lightest chargino $\chi_1^\pm$ and the gluino. The gluino can be either a Majorana particle, as in the MSSM, or a Dirac particle $\tilde{g}_D$, which is equivalently described by the two Majorana mass eigenstates $\tilde{g}_1$ and $\tilde{g}_2$ -- See Appendix for details. We set the masses of the other supersymmetric particles to 10 TeV, effectively decoupling them from the theory. The bino, which is the lightest supersymmetric particle (LSP) and thus stable, is assumed to have a mass of 100 GeV. The chargino mass is set to 150 GeV, but it is only relevant for Sec.~\ref{distinguishing}. We then explore different configurations for the squark and gluino masses.

We use Madgraph~\cite{MG} as the first step to generate parton level Monte-Carlo samples. We introduced in Madgraph the Dirac gluino model using an interface with Feynrules~\cite{Feynrules}. We also use the interface with BRIDGE~\cite{BRIDGE}, to handle long decay chains. The samples are generated at 7 TeV, with the baseline cuts used by the ATLAS analysis~\cite{Aad:2011ib}. We then shower the samples with PYTHIAv6.4~\cite{PYTHIA}, and use a ROOT-based macro to implement some of the detector effects from a detector fast simulator called ATLFAST~\cite{ATLFAST}. Those detector effects include jet clustering and energy-momentum smearing.

%%%%%%%%%%%%%%%%%%%%%%%%%%%%%%%%%%%%
%%%%%%%%%%%%%%%%%%%%%%%%%%%%%%%%%%%%
\section{The Exclusion Limit of Squark and Gluino Masses}
\label{exclusion}

\subsection{ATLAS analysis}
\label{ATLASan}

In this section we briefly describe the analysis done by the ATLAS collaboration in the multijets and missing transverse momentum channels~\cite{Aad:2011ib} with 1.04 fb$^{-1}$ of data. The selection is purely hadronic and a lepton veto is applied. The exclusion limits are shown in a $(m_{\tilde{g}},m_{\tilde{q}})$-plane, where all other supersymmetric particles except the neutralino are decoupled. 

Their strategy is to separate the signal in five bins of different number of jets and missing energy cuts. In all bins, the leading jet has $p_T>$ 130 GeV, and jets with $p_T>$ 40 GeV are used for the counting of number of jets, $n_j$. The variable $M_{eff}$ is built by summing missing energy and the jets characterizing the different signal regions. Additional cuts on the separation between missing momentum and jets, and on the ratio of missing energy to $M_{eff}$ are applied. The signal regions correspond to $n_j \geqslant$ 2 and 3, two more regions with $n_j\geqslant$ 4 and two different cuts on $M_{eff}$, and a final region with at least four jets with $p_T\geqslant$ 80 GeV and $M_{eff}> 1100$ GeV.

\subsection{The exclusion limits for Dirac gluinos}

\begin{figure}[h!]
\centering
\includegraphics[scale=0.25]{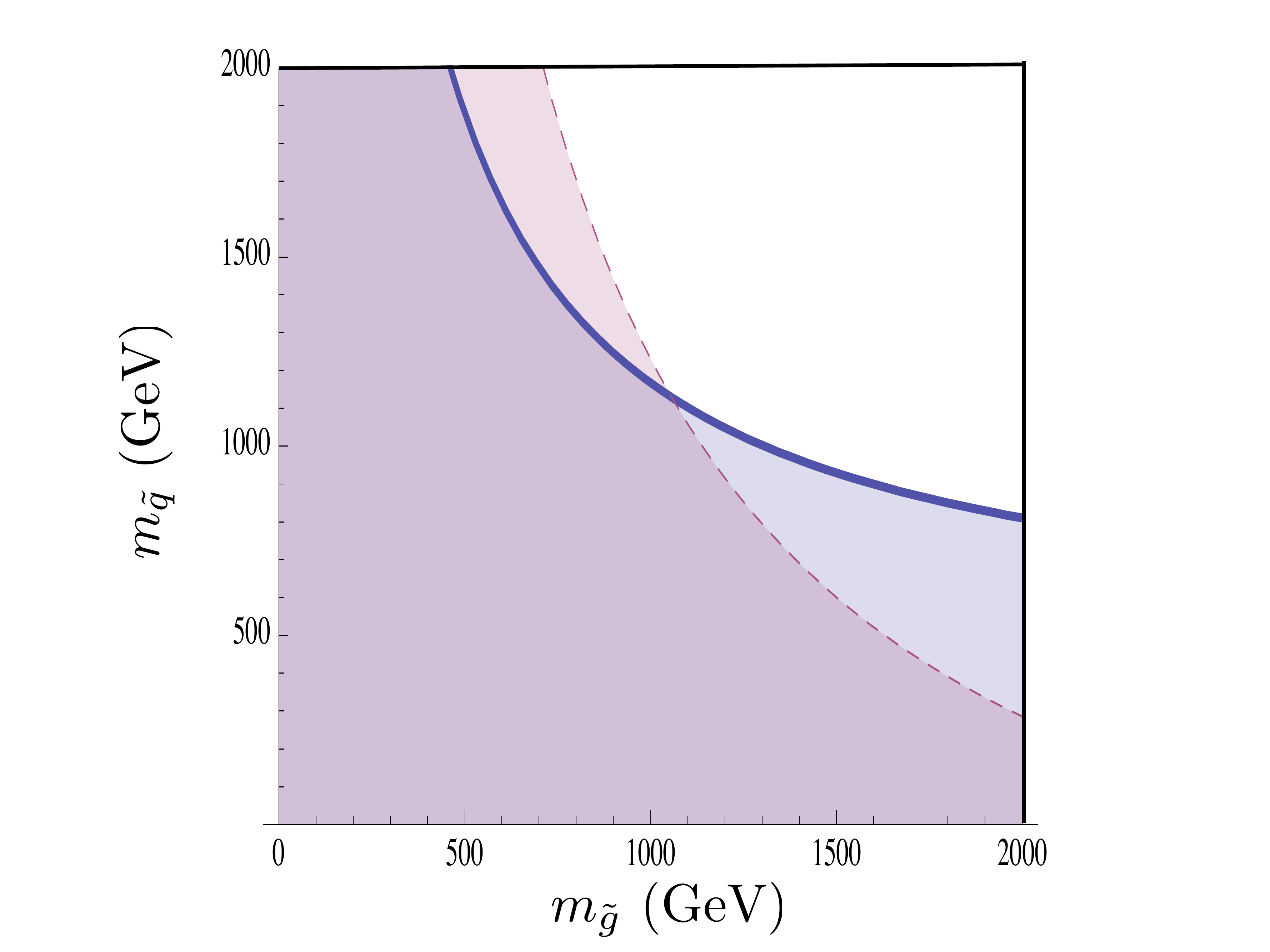}
\caption{The exclusion limit in the
 $(m_{\tilde{g}},m_{\tilde{q}})$-plane in the \None (solid line) and in
 the \Ntwo model (dashed line).}
\label{exclusionlimit}
\end{figure}

As explained in section \ref{general}, the overall production cross section of supersymmetric particles behaves differently in the \None and \Ntwo cases. Since searches are based on the assumption that SUSY is realized as \None, this feature affects the sensitivity of the searches for supersymmetry and the exclusion limits in the
$(m_{\tilde{g}},m_{\tilde{q}})$-plane. The exclusion limit in the MSSM (\None),
as obtained by ATLAS~\cite{Aad:2011ib}, is shown by the solid line in
figure \ref{exclusionlimit}. For the Dirac case this changes, because
the enhanced production cross section of gluino pairs increases the
sensitivity of the search in the low gluino mass region,
and the suppressed squark pair production cross section decreases the
sensitivity in the low squark mass region. Therefore the exclusion limit
becomes tilted, favoring a heavy gluino and leaving more room for a
light squark. The exclusion limit in the Dirac case is shown by the
dashed line in Fig.~\ref{exclusionlimit}. 

We obtain this limit by the
following procedure:
The ATLAS search uses five overlapping bins, divided by number of jets, the
effective mass of the event and the transverse momentum of the jets. For
each point in the $(m_{\tilde{g}},m_{\tilde{q}})$-plane, the bin with
the best sensitivity is used to obtain the exclusion limit. For details,
see ~\cite{Aad:2011ib}. We produced a sample of events with
MadGraph using the same model assumptions as the ATLAS search (see Sec.~\ref{ATLASan}). We then run a simulation as described in Sec.~\ref{simulation}, applied the same cuts as the ATLAS analysis and
selected the bin with the highest number of events as
the bin used for the exclusion. For each point, we then produced a
sample of events in the Dirac model with the same gluino mass, and
varied the squark mass to get the same number of events in the exclusion
bin. The points in the $(m_{\tilde{g}},m_{\tilde{q}})$-plane obtained by
this method were then connected to form the exclusion limit in the Dirac
model.

%%%%%%%%%%%%%%%%%%%%%%%%%%%%%%%%%%%%
%%%%%%%%%%%%%%%%%%%%%%%%%%%%%%%%%%%%
\section{Distinguishing Between Dirac and Majorana Gluinos}
\label{distinguishing}
In the previous section we saw that if the gluino is a Dirac-particle, the exclusion limits set by ATLAS and CMS do not apply as such. There is a notable reduction in the exclusion limit on the squark mass, especially if the gluino is relatively heavy, of the order of 2 TeV\footnote{Note that, as the gluino becomes heavier, the difference between the Majorana and Dirac cases is reduced. But at moderate values of the gluino mass around 2 TeV, the effect of a t-channel exchange of the gluino is still dominant.}. Thus, we have shown that in \Ntwo SUSY it is still possible that first and second generation squarks exist at or below the TeV-scale. In this section we turn our focus to the question of identifying the Dirac or Majorana nature of the gluino, provided experimental evidence of the existence of supersymmetric particles is found\footnote{ For a different proposal of distinguishing \None and \Ntwo SUSY at colliders, see Ref.~\cite{Choi:2010gc}.}.

In section \ref{general} we described the three parton level final states relevant for the production of supersymmetric particles. The squarks and gluinos then cascade decay to Standard Model particles and the LSP. If one imposes a lepton veto, the final states we are looking at are multiple jets and missing transverse
momentum such as $\tilde{q}\tilde{q}\rightarrow 2j+\ETm$, $\tilde{q}\tilde{g}\rightarrow 3j+\ETm$ and  $\tilde{g}\tilde{g}\rightarrow 4j+\ETm$.

However, by imposing a lepton veto we miss other decay chains which tell us about the chirality of the squark. Indeed, while the right-handed squarks decay almost exclusively to a quark and the LSP, the left-handed
squarks have a large branching ratio to $\tilde{q}_L\rightarrow q\chi_1^\pm$, where the
chargino then decays into a $W$ plus the LSP. The $W$ can then decay hadronically or leptonically. Hence the detector level final state depends on the chirality of the produced squarks, with left-handed squarks producing a higher jet and/or lepton multiplicity.

This feature can be used to distinguish between \None and \Ntwo. In the MSSM (\None case) the two jets final state mostly
originates from the production of two right-handed squarks, but in the
Dirac scenario this process is forbidden. Therefore the two jets final state
is only accessible via the production of a right handed squark and a
right handed antisquark, which requires an antiquark in the initial
state and is thus suppressed by the small antiquark PDFs, or through a
subdominant decay of a left handed squark in the
$\tilde{q}_L\tilde{q}^\prime_R$ final state. Therefore the two jets final
state is suppressed in the Dirac case compared to the MSSM. On the other
hand the gluino pair production process, which leads to final states
with a high jet multiplicity, is enhanced in the Dirac model. Thus we get
more events with $n_j\geq4$ compared to the MSSM. The overall effect is that the Dirac case tends to produce more events with a high jet multiplicity and less events with a low multiplicity than the Majorana case.  This is also apparent in table \ref{atlasbins_table}, which shows some of the results of our simulation. The number shown for each bin is the ratio of production cross section of squarks and gluinos times the acceptance of each bin for the \None and \Ntwo cases. The bin with the highest acceptance for each value of squark and gluino masses is highlighted with a bolded (underlined) font in the \None (\Ntwo) case. The cross sections aree based on a tree level calculation performed with MadGraph, and the overall scale would have to be adjusted by a computation at NLO, but this $K$-factor is common to \None and \Ntwo and therefore it factors out in the ratio of events.  One can see a clear tendency towards the higher multiplicity bins in the Dirac model. The ratio of the number of events in the $2j+\ETm$ final state and the number of events
with four or more jets plus missing transvere momentum can then be used
to separate the Dirac scenario from MSSM. A low number for this ratio is
characteristic to the Dirac model.

\begin{table}
\begin{tabular}{|c|c|c|c|c|c|c|}
\hline
$m_{\tilde{g}}$ [GeV]&$m_{\tilde{q}}$ [GeV]& bin 1 & bin 2 & bin 3 & bin 4 & bin 5 \\ \hline
1100 & 1500 & 0.49 & 3.03 & \textbf{\underline{2.92}} & 2.90 & 2.87 \\ \hline
1300 & 1200 & 0.14 & \textbf{0.33} & \underline{0.44} & 0.45 & 0.43 \\ \hline
1500 & 1250 & 0.075 & \textbf{\underline{0.23}} & 0.26 & 0.26 & 0.29 \\ \hline
1500 & 1700 & 0.086 & 0.30 & \textbf{\underline{0.42}} & 0.42 & 0.45 \\ \hline
1800 & 700 & 0.090 & \textbf{\underline{0.16}} & 0.13 & 0.15 & 0.19 \\ \hline
1800 & 2000 & 0.058 & 0.19 & \textbf{\underline{0.28}} & 0.28 & 0.34 \\ \hline
\end{tabular}
\caption{The ratio of the number of events in \None/\Ntwo SUSY models, for the bins used in the ATLAS search for hadronic events and missing transverse momentum~\cite{Aad:2011ib}.}
\label{atlasbins_table}
\end{table}

To obtain the best current limits, we have followed the ATLAS search which assumed a lepton veto. However, for the purpose of distinguishing between \None and \Ntwo, the number of events with one or more leptons compared to the number of events without leptons is a useful observable. If we allow for one or more leptons in the two jets plus missing transverse momentum bin, the $\tilde{q}_L\tilde{q}_R$ final state may end up in this bin via the decay $\tilde{q}_L\rightarrow q\chi_1^\pm \rightarrow ql^\pm \nu_l \chi^0$. Since the $\tilde{q}_R\tilde{q}^\prime_R$  partonic level final state is not accessible in the Dirac model, this decay chain makes up for a notable part of the two jets events. Therefore the ratio of events in the two jets bin that contain one or more leptons to the number of events in that bin that do not contain isolated leptons is high in the Dirac model. In \None SUSY the majority of events in the two jets bin is produced via the $\tilde{q}_R\tilde{q}^\prime_R$ partonic level final state. Therefore allowing for one or more leptons does not add significantly to the total number of events in that bin, since the leptonic decay of the left handed squark in the $\tilde{q}_L\tilde{q}_R$ final state is a subdominant process compared to the hadronic decay. Thus a good separation between the models is achieved by comparing the ratio of events in the two jets plus missing transverse momentum bin that do or do not contain one or more leptons. This ratio is higher in the Dirac case compared to the Majorana case.

\begin{figure}[t]
\centering
\includegraphics[scale=0.3]{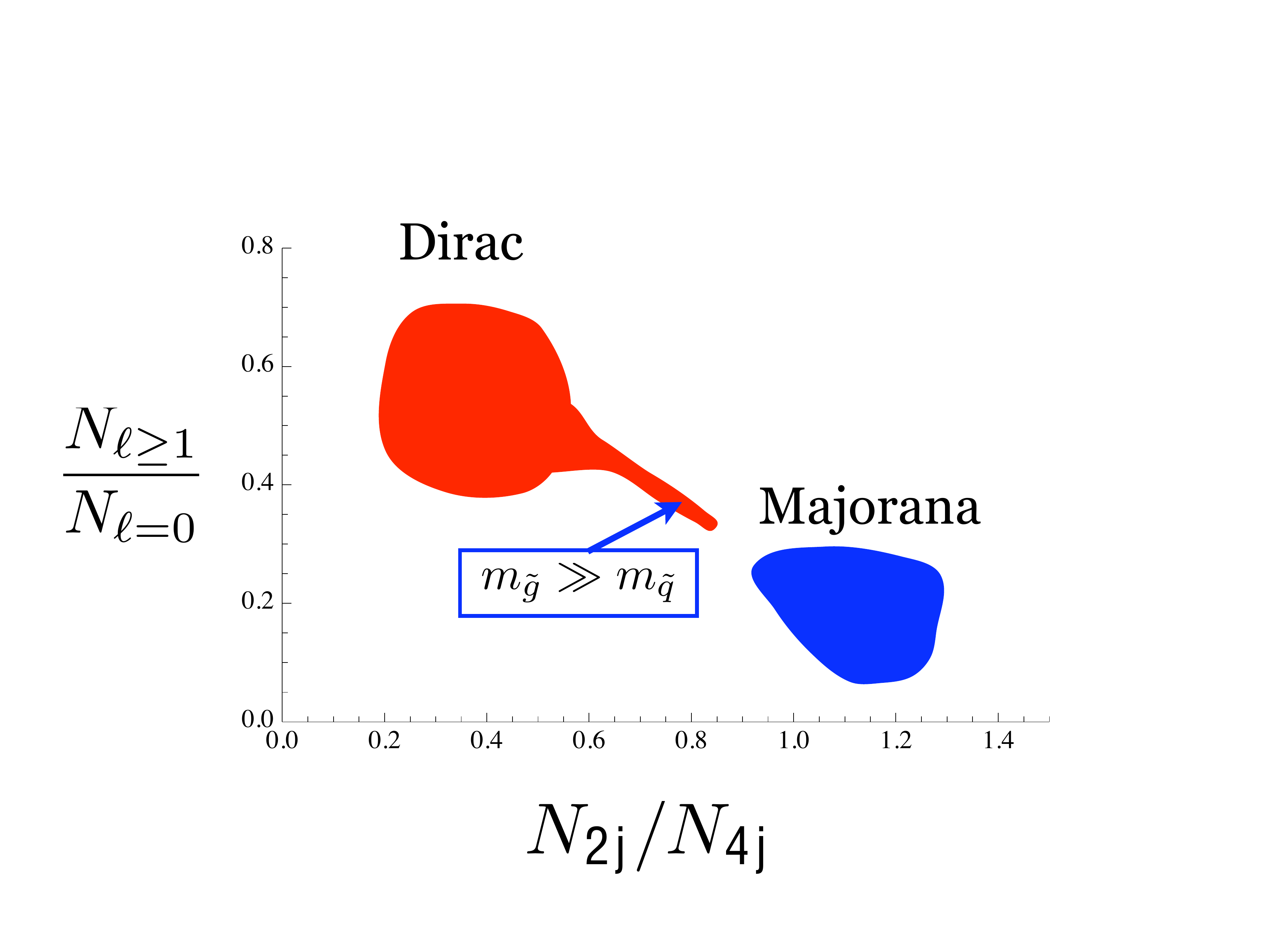}
\caption{A schematic illustration of the two observables used for distinguishing between the Dirac model and the MSSM. On the $x$-axis is the ratio of the number of events with two jets and the number of events with four or more jets. On the $y$-axis is the ratio of events with or without leptons in the two jets bin. The two areas shown are obtained by varying the squark and gluino masses in the two models. When the gluino is much heavier than the squark,  both \None and \Ntwo cases become indistinguishable.}
\label{leptons_jets}
\end{figure}

The best separation between \None and \Ntwo is achieved by combining these two observables. This is schematically shown in figure \ref{leptons_jets}, where the ratio of events in the two jets and four jets bins is plotted in the $x$-axis and the ratio of events with or without leptons in the two jets bin is shown in the $y$-axis. We generated event samples with different mass configurations, but all obeying $m_{\tilde{q}}<m_{\tilde{g}}$ (see discussion below for the $m_{\tilde{q}}>m_{\tilde{g}}$ case). We then placed the resulting point of each mass configuration on the $(x,y)$-plane. The points corresponding to \None end up on the lower right corner of the plane as expected, and the points in \Ntwo populate the upper left corner of the plane. When the gluino mass becomes very heavy, however, the gluino starts to decouple from the theory. Then the squark pair production process begins to be dominated by the $s$-channel gluon exchange diagram, which contains no information of the Dirac- or Majorana-nature of the gluino. Therefore the points corresponding to the Dirac model begin to approach the MSSM part of the $(x,y)$-plane in the limit where the gluino mass is very heavy. This is also illustrated in table \ref{leptons_table}, where the values of these observables is shown for two points. One of the points is in the area where the gluino is not too heavy and the separation between \Ntwo and \None is good, and the other one shows how the separation gets worse as the gluino mass grows and the squark mass gets small. We conclude that, except for the case where the gluino is extremely heavy and thus practically decoupled from the theory, the two observables plotted in the $x$ and $y$ axes of the plane can be effectively used to identify if the gluino is a Dirac- or a Majorana-particle.

\begin{table}
\begin{tabular}{|c|c|c|c|c|c|c|}
\hline
$m_{\tilde{g}}$ [GeV]&$m_{\tilde{q}}$ [GeV]&\begin{tabular}{l}$n_j=2$\\$n_l=0$\end{tabular}&\begin{tabular}{l}$n_j=2$\\$n_l\geq 1$\end{tabular}&\begin{tabular}{l}$n_j\geq 4$\\$n_l=0$ \end{tabular}&$\frac{n_l\geq1}{n_l=0}$&$\frac{n_j=2}{n_j=4}$ \\ \hline
\textbf{\None} & ~ & ~ & ~ & ~ & ~ & ~ \\ \hline
1500 & 1250 & 0.57 & 0.11 & 0.48 & 0.19 & 1.19 \\ \hline
1800 & 700 & 15.5 & 2.34 & 12.5 & 0.15 & 1.24 \\ \hline
\textbf{\Ntwo} & ~ & ~ & ~ & ~ & ~ & ~ \\ \hline
1500 & 1250 & 0.043 & 0.028 & 0.13 & 0.65 & 0.34 \\ \hline
1800 & 700 & 1.39 & 0.41 & 1.61 & 0.29 & 0.86 \\ \hline
\end{tabular}
\caption{The relevant observables used for the identification of the Dirac nature of the gluino: the number of events with two jets versus the number of event with four or more jets, and the number of events with one or more leptons and two jets versus two jets and no leptons, in the \None and \Ntwo cases for two points in the $(m_{\tilde{g}},m_{\tilde{q}})$ plane.}
\label{leptons_table}
\end{table}

The above picture is less clear if the gluino is lighter than the
light squarks, because then all the squarks decay dominantly to a
quark and a gluino, which leads to higher jet multiplicities also in the \None case.   The lepton vs no-lepton discriminant is still a good discriminant between the two scenarios, but  the ratio of two jets finals states and four or more jets
final states is very small in both models in this region of the
parameter space, and one would have to look into observables with two more jets, namely the ratio of four to six jets versus the ratio of lepton and no lepton in the four jet final state.

Let us finish by mentioning the option of pseudo-Dirac gluinos, i.e., Dirac gluinos with a small Majorana admixture. A small splitting between the two Majorana states in the Dirac gluino would not affect the results in this section, as they are based on multijet final states with high-$p_T$ requirements. 
 
%%%%%%%%%%%%%%%%%%%%%%%%%%%%%%%%%%%%
%%%%%%%%%%%%%%%%%%%%%%%%%%%%%%%%%%%%
\section{Conclusions}

In this paper we have shown that the current bounds on Supersymmetry are drastically altered when one considers the possibility that Supersymmetry could be realized as an \Ntwo version in the gauge sector. This is a key question nowadays, as the LHC is pushing the bounds on Supersymmetry to uncomfortable areas for natural solutions of the hierarchy problem.

We have taken the best current limits from ATLAS and redone their analysis --within the tools available to theorists-- to estimate the limit on \Ntwo Supersymmetry. We find that, in \Ntwo supersymmetry, squarks as light as few hundreds of GeVs are still allowed. This study illustrates the following important point, that experimental studies of Supersymmetry are remarkably dependent on the assumption about the number of supersymmetries.

We have gone a step ahead, and found a suitable combination of observables which, provided an excess in multijet events is found, can be used as a discriminant between \None and \Ntwo supersymmetry. Since natural models of supersymmetry are realizations of either \None or \Ntwo in the gauge sector, this discriminant is effectively counting the number of supersymmetries in Nature.

\section*{Acknowledgements}

We would like to thank Randy Lewis for collaboration in the initial stages of this work, and for his support thorough the project. VS and MH are partly supported by an NSERC Discovery grant, and MK was supported by a summer NSERC USRA.

\section*{Appendix}

Here we will show in detail how the \Ntwo gluino sector can be expressed in terms of one gluino which has a Dirac mass term. We will start by defining some notation and conventions, then we show how the Lagrangian is transformend from the flavor basis to the mass basis, expressed in terms of the two Majorana mass eigenstates. Finally we show that this is equivalent to the Lagrangian of one Dirac particle.

We use the chiral basis, and the notation in~\cite{Martin:1997ns}. 
We define two-component Weyl spinors
with dotted and undotted indices: $\eta_\alpha$ and $\eta_\alphadot$, as
complex, anticommuting objects with $\alpha=1,2$ and
$\alphadot=1,2$. The spinors with dotted and undotted indices are
related via Hermitian conjugation:

\be
\eta_\alphadot^\dagger:=(\eta_\alpha)^\dagger=(\eta^\dagger)_\alphadot,
\qquad \eta^{\dagger\alphadot}=(\eta^\alpha)^\dagger.
\ee

The spinor indices are raised and lowered using the antisymmetric symbol
$\epsilon^{12}=-\epsilon^{21}=-\epsilon_{12}=\epsilon_{21}=1$, 
$\epsilon_{ii}=\epsilon^{ii}=0$:
\be
\eta_\alpha=\epsilon_{\alpha\beta}\eta^\beta, \quad \eta^\alpha=\epsilon^{\alpha\beta}\eta_\beta,
\ee
and similarly for the spinors with dotted indices. In the following, we will suppress contracted indices. 

A four-component spinor is written in terms of the two-component
spinors as
\be
\Psi=\begin{pmatrix} \eta_\alpha \\ \chi^{\dagger\alphadot} \end{pmatrix},
\ee
and thus its conjugate spinor is
\be
\bar{\Psi}=\Psi^\dagger\gamma^0=\begin{pmatrix} \eta^\dagger_\alphadot &
				 \chi^\alpha \end{pmatrix} \gamma^0=
				 \begin{pmatrix} \chi^\alpha &
				  \eta^\dagger_\alphadot \end{pmatrix}.
\ee
In our conventions, the left and right handed fields are then simply the upper and lower components of the four-component field:
\be
\Psi_L=P_L\Psi=\begin{pmatrix} \eta_\alpha \\ 0 \end{pmatrix}, \quad \Psi_R = P_R\Psi=\begin{pmatrix} 0 \\ \chi^{\dagger\alphadot} \end{pmatrix}.
\ee
For SM Dirac fields, such as a quark field $Q$, we therefore write
\be
Q=\begin{pmatrix} q_L \\ q_R \end{pmatrix}, \quad Q_L=\begin{pmatrix} q_L \\ 0 \end{pmatrix}, \quad Q_R=\begin{pmatrix} 0 \\ q_R \end{pmatrix},
\ee
and
\be
\bar{Q}=\begin{pmatrix} q^\dagger_R & q^\dagger_L \end{pmatrix}, \quad
		    \overline{Q_L}=\begin{pmatrix} 0 & q^\dagger_L
			      \end{pmatrix}, \quad
						  \overline{Q_R}=\begin{pmatrix}
							    q^\dagger_R
							    & 0
							    \end{pmatrix}.
\ee

The charge conjugate of a 4-component spinor is
\be
\Psi^c=C\bar{\Psi}^T=\begin{pmatrix} -i\sigma^2\chi^\alpha \\ i\sigma^2\eta^\dagger_\alphadot \end{pmatrix} = \begin{pmatrix} \epsilon_{\alpha\beta}\chi^\beta \\ \epsilon^{\alphadot\betadot}\eta^\dagger_\betadot \end{pmatrix}= \begin{pmatrix} \chi_\alpha \\ \eta^{\dagger\alphadot} \end{pmatrix}.
\ee
Majorana particles must obey $\Psi^c=\Psi$, implying the Majorana
condition $\chi=\eta$. 

Let us move into the gluino mixing. Assuming we have two Majorana gluinos:
\be
\go=\bmat \eta_\alpha \\ \eta^{\dagger\alphadot} \emat, \quad \gop=\bmat
{\etap}_\alpha \\ {\etap}^{\dagger\alphadot} \emat.
\ee
The mass term in the general case is
\be
\mathcal{L}_m=-\frac12\left(m^\prime_3\bar{\go}^\prime\gop+m_3\bar\go\go+m_D(\bar{\go}^\prime\go+\bar{\go}\gop)\right).
\ee
We will be interested in the Dirac limit, i.e. $m^\prime_3=m_3=0$. In
terms of the two-component spinors the mass Lagrangian is then
\be
\mathcal{L}_m=-\frac12m_D(\bar{\go}^{\prime}\go+\bar{\go}\gop)=-m_D(\etap\eta+\etap^\dagger\eta^\dagger).
\ee
The mass eigenstates are given by
\bea
\go_1&=&\frac{1}{\sqrt{2}}\bmat \eta_\alpha+{\etap}_\alpha \\
\eta^{\dagger\alphadot}+{\etap}^{\dagger\alphadot} \emat =\bmat
\chi_\alpha \\ \chi^{\dagger\alphadot}\emat, \\ 
\go_2 &=& \frac{i}{\sqrt{2}}\bmat {\etap}_\alpha-\eta_\alpha \\
\eta^{\dagger\alphadot}-{\etap}^{\dagger\alphadot} \emat=\bmat
\chip_\alpha \\ \chip^{\dagger\alphadot}  \emat.
\eea
These are Majorana particles, as $\go_1^c=\go_1$ and $\go_2^c=\go_2$.  One can easily check that the mass Lagrangian is indeed diagonal in this basis.

We will now examine the original Lagrangian term by term, transforming
into the mass eigenbasis. First, the kinetic term:
\be
\begin{split}
\mathcal{L}_K&=\frac{i}{2}(\bar{\go}\slashed{\partial}\go+\bar{\go}^{\prime}\slashed{\partial}\gop) \\
&=\frac{i}{2}(\eta\sigma^\mu\partial_\mu\eta^\dagger+\eta^\dagger\bar\sigma^\mu\partial_\mu\eta+\etap\sigma^\mu\partial_\mu\etap^\dagger+\etap^\dagger\bar\sigma^\mu\partial_\mu\etap) \\
&=\frac{i}{2}(\chi\sigma^\mu\partial_\mu\chi^\dagger+\chi^\dagger\bar\sigma^\mu\partial_\mu\chi+\chip\sigma^\mu\partial_\mu\chip^\dagger+\chip^\dagger\bar\sigma^\mu\partial_\mu\chip) \\
&=\frac{i}{2}(\bar{\go}_1\slashed{\partial}\go_1+\bar{\go}_2\slashed{\partial}\go_2).
\end{split}
\ee
Similarly, the gauge term is
\be
\begin{split}
\mathcal{L}_{\rm
 gauge}&=-ig_sf^{ijk}(\bar{\go}^i\slashed{G}^j\go^k+\bar{\go}^{\prime i}\slashed{G}^j\go^{\prime k}) \\
&=-ig_sf^{ijk}(\bar{\go}_1^i\slashed{G}^j\go_1^k+\bar{\go}_2^i\slashed{G}^j\go_2^k).
\end{split}
\ee
 
The degrees of freedom present in $\go_1$ and $\go_2$ can be combined into
one Dirac spinor $\go_D$, given by
\be
\go_D=\frac{1}{\sqrt{2}}(\go_1-i\go_2)=\bmat \etap_\alpha \\
\eta^{\dagger\alphadot} \emat.
\ee

In terms of the Dirac gluino, the mass term is
\be
\mathcal{L}_m=-m_D(\eta\etap+\eta^\dagger\etap^\dagger)=-m_D\bar{\go}_D\go_D,
\ee
as expected. The gauge term and the Kinetic term are
\be
\begin{split}
\mathcal{L}_{\rm
 gauge}&=-ig_sf^{ijk}\left(\bar{\go}_D^i\slashed{G}^j\go_D^k+\bar{\go}_D^{c\ i}\slashed{G}^j\go_D^{c\
 k}\right), \\
\mathcal{L}_K&=\frac{i}{2}\left(\bar{\go}_D\slashed{\partial}\go_D+\bar{\go}_D^c\slashed{\partial}\go_D^c\right).
\end{split}
\ee
The trilinear quark-squark-gluino coupling term is
\be
\mathcal{L}_{q\go\sq}^{(1)}=-g_s\left(\overline{Q_L}\go_D\sq_L-\overline{Q_R}\go_D^c\sq_R\right)+h.c.\ .
\ee
Thus we have seen that the gluino sector is equivalently described either by two degenerate Majorana particles $\go_1$ and $\go_2$ or by one Dirac particle $\go_D$. To validate our Monte Carlo implementation, we implemented the \Ntwo model twice, using both descriptions, and checked that the numerical results agree between these models. We also checked that decoupling one of the Majorana gluinos reduces the \Ntwo model back to MSSM, and our numerical results confirm that.

%%%%%%%%%%%%%%%%%%%

\end{document}